\documentstyle{aipproc}

\begin{document}
\title{Pulsar Counterparts of Gamma-Ray Sources}

\author{ P.A.~Caraveo$^1$ and G.F.~Bignami $^{2,1}$}
\address{ $^1$Istituto di Fisica Cosmica del CNR, Via Bassini, 15, 20133 Milano, 
Italy\\
$^2$Agenzia Spaziale Italiana, Via di Villa Patrizi 13, Roma, Italy}

%\lefthead{LEFT head}
%\righthead{RIGHT head}
\maketitle

\begin{abstract}
The EGRET catalogue of unidentified X-ray sources has more objects along the
galactic disk than at  high galactic latitude, where identifications are
comparatively easier. On the other hand, the Egret/GRO  mission has already
identified several known radio pulsars as gamma-ray sources as well as
discovering  Geminga's nature as a pulsar. If Geminga is not a unique case, as
it is very likely not to be, than other  galactic sources could, in fact, be
radio quiet isolated neutron stars. \\For these, the identification work is
extremely difficult and should anyway start from high resolution X-ray/optical
data.

\end{abstract}

\section*{Introduction}

Isolated Neutron stars (INSs) are the only galactic objects surely identified
as $\gamma$-ray emitters. Although not completely understood, as far as the
emission mechanism is concerned, their phenomenology is fairly well known. \\
Combining multiwavelength observations spanning the entire electromagnetic
spectrum, we have learned that diversity is the rule amongst $\gamma$-ray
emitting INSs. With the notable exeption of the Crab, light curves are
remarkably different in separate energy ranges (see e.g. Kanbach, 1997, for a
review), hinting that alternate beaming geometries are possibly associated to
different emission mechanisms. Moreover, the fraction of the rotational energy
loss re-emitted in  high-energy $\gamma$-rays ranges from a fraction of a
percent, in the case of the Crab, to the quasi-totality for PSR1055-52 (see
e.g. Goldoni et al, 1995). This does not translate directly into luminosity
because the rotational energy loss of young pulsars is orders of magnitude
larger than that of older objects, so that, in spite of lower efficiency, young
pulsars are indeed the brightest $\gamma$-ray sources in our Galaxy. \\ With six
successful identifications (counting also PSR 1915+32, which is not included in
the $2^{nd}$ EGRET catalogue since it is detected only as a pulsating source)
amongst the 45 low latitude sources seen by EGRET (Thompson et al, 1995,
Thompson et al, 1996), it is natural to explore the possibility that at least a
fraction of the remaining low latitude sources belong to the same class of
compact objects.  Indeed, many searches for pulsars inside COS-B and EGRET
error boxes have been and are being carried out: all in all pulsars have
received far more attention than any other galactic population thought to be a
possible source of high-energy  $\gamma$-rays.  \\ Why is the pulsar hypothesis
so successful amongst $\gamma$-ray astronomers? \\ In a branch of astronomy 
hampered by poor angular resolution and low counting rates, a pulsar
identification is by far the most unambiguous one. When the light curve
obtained folding the $\gamma$-ray photons at a known pulsar period is
statistically compelling, one should not worry about chance superposition nor
about lengthy follow-up observations: the source is identified for sure. \\ This
is the appeal of a pulsar identification and this is why pulsar searches have
been performed on $\gamma$-ray error boxes as soon as  COS-B discovered the
UGOs (Unidentified Gamma Objects, see Bignami and Hermsen, 1983 for a review).
The majority of the searches zeroed in on Geminga, but a fair number of the
sources of the $2^{nd}$ COS-B catalogue were surveyed with no luck. However,
one of the COS-B UGOs turned out to be a pulsar which was discovered in a
routine survey for southern pulsars ten years after the end of the COS-B
mission. It is the case of 2CG342-02, identified in 1992 with PSR 1706-44
(Thompson et al, 1992), a Vela-like pulsar at 2 kpc.  \\ COS-B had two, with PSR
1706-44 three, radio pulsars amongst 22 low latitude sources. EGRET has five
(six if we count the tentative identification od PSR 0656+14 and seven if we
consider also PSR 0540-69, seen by all CGRO instruments but EGRET) amongst 45
sources. In spite of the increased sensitivity of EGRET,  the ratio between
pulsar identification and total number of sources appears to be constant. Also
the lack of results, experienced at the time of COS-B, appears to be
unchanged.   Dedicated radio searches (Nice and Sayer, 1997), aimed precisely
at the search for radio pulsars inside  the error boxes of 10 of the brightest
EGRET sources, yielded null results, showing that the straightforward radio
pulsar identification is not the only possible solution to the enigma of the
unidentified high-energy $\gamma$-ray sources. This is further strengthned by
the work of Nel et al. (1997) who investigated 350 known pulsars finding few
positional coincidences but no significant timing signature for any of the
pulsars in the survey.

\section*{A different Approach}
Indeed, $\gamma$-ray astronomy does offer a remarkable example of an Isolated
Neutron Star (INS) which behaves as a pulsar as far as X-and-$\gamma$ astronomy
are concerned but has little, if at all, radio emission.  As an established
representative of the non-radio-loud INSs (see Caraveo, Bignami and Tr\"umper,
1996 for a review), Geminga offers a more elusive template behaviour: prominent
in high energy $\gamma$-rays, uneventful in X-rays and downright faint in
optical, with sporadic or no radio emission. Although the latitude distribution
of unidentified EGRET sources shows that, in average, they are at least 10
times as distant as Geminga (Mukherjee et al, 1995), the multiwavelength
behaviour of this source is hard to beat when one tries to link gamma-ray
sources to compact objects in the absence of a radio signal. While one should
always keep a totally open mind and be ready to find something new and
different, Geminga is the template observers have in mind when planning
observing strategies. Unfortunately, in spite of our knowledge of the behaviour
of the real Geminga, the study of a Geminga-like source still represents a
great challenge to observers since it defies well established techniques. To
appreciate such a challenge let us briefly review the many steps that lead to
the identification and the understanding of this object with an aim to find the
signature to look for.

\section*{The Many Firsts of Geminga}
Briefly, the source was discovered in high energy $\gamma$-ray  by the SAS-2
satellite in 1972 (Fichtel et al, 1975), an X-ray  counterpart, suggesting
position and distance, has been proposed  in 1983 (Bignami et al. 1983)  and an
optical one, refining the  position, in 1987/88 (Bignami et al, 1987, Halpern
and Tytler,  1988). However, the breakthrough came with the discovery of the 
237 msec periodicity in the ROSAT data (Halpern and Holt, 1992).  Finding the
same periodicity in the simultaneous high energy  $\gamma$-ray data of the
EGRET instrument (Bertch et al, 1992), as  well as in the old archival COS-B
(Bignami and Caraveo,  1992) and SAS-2 data (Mattox et al, 1992), yielded the
value  of the period derivative and thus of the  object's energetics. The
discovery of the proper motion of the  proposed optical counterpart (Bignami et
al, 1993) confirmed the  optical identification and, thus, provided the
absolute positioning  of Geminga to within the systematic uncertainty of the 
Guide Star Catalogue, i.e. 1". Next came the measure of  the source parallactic
displacement, yielding a precise measure of  its distance (Caraveo et al,
1996). \\ More HST observations, confirming and refining difficult measurements
with ground-based instruments, have shown that a broad feature, centered at
$\lambda=5998 \AA$ and with a width of 1,300 $\AA$, is superimposed to the
Rayleigh-Jeans continuum, as extrapolated from the soft X-rays (Bignami et al,
1996; Bignami, 1997). If interpreted as an ion-cyclotron emission, it implies,
for Z/A=1, a B field of $3.25~10^{11}$ Gauss as opposed to the value of
$1.5~10^{12}$ obtained, theoretically, using the Period and Period derivative.
This is the first time that the magnetic field of a neutron star is directly
measured. Recently, the phenomenology of the source at high energies has been
considerably enriched, owing to the very precise positioning of the optical
counterpart. The possibility to link HST data to the Hipparcos reference frame,
yielded the position of Geminga to an accuracy of 0.040 arcsec, a value unheard
of for the optical position of a pulsar, or of an object this faint (Caraveo et
al, 1997). This positional accuracy has allowed to phase together data
collected over more than 20 years by SAS-2, COS-B and EGRET, unveiling very
promising timing residuals (Mattox et al, 1997). The many "firsts" of Geminga
are summarized in Table 1. \\ Quite surprisingly, some of the key parameters of
Geminga are now known with an accuracy better than available for the Crab
pulsar. This is due in part to the 20 year long chase (see Bignami and Caraveo,
1996 for a review), in part to the remarkable stability of this object which
rendered possible to phase together such a long time span of $\gamma$-ray data.

\begin{table*}
\caption{GEMINGA(1973-present)}
\begin{tabular} {|l|} \hline 

$\bullet$ $1^{st}$~unidentified~$\gamma$-ray~source \\ \hline

$\bullet$ $1^{st}$~INS~discovered~through~high-energy~emission~and~identified~through~its~X~and~$\gamma$-rays\\

$\bullet$ $1^{st}$~INS~identified~without~the~help~of~radio~astronomy \\ \hline

$\bullet$ $1^{st}$~INS~optically~identified~through~its~proper~motion \\

$\bullet$ $1^{st}$~INS~the~distance~of~which~is~measured~through~its~
optical~parallax\\ \hline

$\bullet$ $1^{st}$~direct~view~~in~optical/UV~of~the~surface/photosphere
 of~a~NS \\

$\bullet$ $1^{st}$~evidence~for~an~atmosphere~surrounding~NS~crust \\

$\bullet$ $1^{st}$~direct~measurement~of~the~surface~magnetic~field~of~an~INS \\
\hline

$\bullet$ $1^{st}$~INS~the~timing~parameters~of~which~are~determined~
solely~by~high~energy~$\gamma$-ray~data \\ 

$\bullet$ $1^{st}$~optical~measurement~of~absolute~position~of~an~INS~
within~40~mas \\~~~~(This~leads~to~the~first~measurement~of~
the~braking~index~of~a~$10^{5}$~y~old~NS) \\ \hline

$\bullet$ $1^{st}$~evidence~(together~with~PSR0656+14~and~PSR~1055-58)
~of~an~INS~with~a \\~~~~two-component~X-ray~emission~ \\ \hline

\end{tabular}
\end{table*}

\section*{A Strategy for the Future}
There is no question that a sizeble fraction of the EGRET UGOs are galactic. Of
these. It is reasonable to expect at least several to be radio-quiet INSs,
since  no further radio pulsars can be identified. Also, no other compact (or
star-like) class of galactic objects has yet been identified with certainty as
a $\gamma$-ray emitter. It make sense, therefore, to single out further
$\gamma$-ray INSs even if the process might be difficult and tiresome, as in
the case of Geminga. The potential reward will obviously be a better
understanding of the radio-quiet, $\gamma$-ray loud INSs as a class. This would
yield a precious addition to the general neutron star scanty phenomenology.
\\Furthermore, a dedicated search process like that required to nail down an
INS might well yield, as a bonus, new identifications of $\gamma$-ray objects
of serendipitous nature.  \\To plan a strategy for such a search it is as easy
as it is difficult to predict, with any confidence, its probability of success.
This is to say that the only obvious way forward is one similar to the "Geminga
chase" (Bignami and Caraveo, 1996). At the same time, we know from the start
that, although possible Geminga-like, the majority of the UGOs must differ
significantly from Geminga itself. The main difference will be in the absolute
value of their $\gamma$-ray luminosity: objects as (relatively) faint as
Geminga, will never be seen at the distances (several kpc) that UGOs must have
to show their narrow latitude distribution. This could be due to differences in
ages, which in term would yield different ratios of the power emitted in
thermal versus non-thermal processes, both in optical and X-rays. This will
impact on INS visibility. \\Nevertherless, until  a better one is found, the
UGO identification strategy can only be as follows: \\ 1- map the UGO boxes with
X-ray imaging devices \\2- select those few sources which have a very high
$F_{x}/F_{v}$  \\3- search for possible X-ray pulsations \\4- go for optical
IDs, using all possible methods, not forgetting proper motion. \\With respect
to the Geminga chase, carried out in the '80s and early '90s, steps 1 and 4 now
benefit of a far deeper penetration in the sky owing to existing (and upcoming)
orbiting and ground based telescopes. In the next couple of years, for example,
the EPIC instrument on ESA's XMM will be operational as will be (at least) UT1
of ESO's VLT. Their joint usage, if well thought out, will improve by at least
three to four magnitudes on the ROSAT/NTT combination, which did much of the
work on Geminga. The extremely accurate relative astrometry, possible with the
new instruments on HST will detect the much smaller proper motions (and
parallaxes) of more distant INSs. \\It should be exciting to see, within the
next 5 years, how many more Geminga will be seen and if indeed a new different
galactic population, other than INSs, is needed to explain EGRET's UGOs.

\end{document}